\documentclass[aps,prd,reprint,a4paper,showpacs,showkeys,superscriptaddress]{revtex4-1}
\ifx\pdfoutput\undefined
\usepackage[usenames]{color} 
\usepackage[dvips]{graphicx}
\else
\usepackage{color} 
\definecolor{BrickRed}{rgb}{0.85,0.15,0.25}
\definecolor{MidnightBlue}{rgb}{0,0.45,0.85}
\definecolor{ForestGreen}{rgb}{0,0.85,0.45}
\usepackage{graphicx}
\usepackage{epstopdf}
\fi
\usepackage{subfigure}
\usepackage{latexsym,amsmath,amssymb}
\allowdisplaybreaks

\usepackage{umoline}
\newsavebox\CBox

\begin{document}


\title{A Quantal Tolman Temperature}

\author{Yongwan Gim}%
\email[]{yongwan89@sogang.ac.kr}%
\affiliation{Department of Physics, Sogang University, Seoul 121-742,
  South Korea}%

\author{Wontae Kim}%
\email[]{wtkim@sogang.ac.kr}%
\affiliation{Department of Physics, Sogang University, Seoul 121-742,
  South Korea}%

\date{\today}

\begin{abstract}
The conventional Tolman temperature based on the assumption of the traceless condition of energy-momentum
tensor for matter fields is infinite at the horizon if Hawking radiation is involved. However, we note that
the temperature associated with Hawking radiation is of relevance to the trace anomaly,
which means that the traceless condition should be released.
So, a trace anomaly-induced
Stefan-Boltzmann law
is newly derived by employing the first law of thermodynamics and the property of
the temperature independence of the trace anomaly.
Then, the Tolman temperature
is quantum-mechanically generalized according to the anomaly-induced Stefan-Boltzmann law.
In an exactly soluble model, we show that the Tolman factor
does not appear in the generalized Tolman temperature which is eventually finite everywhere,
in particular, vanishing at the horizon. It turns out that the equivalence principle survives
at the horizon with the help of the quantum principle, and some puzzles related to the Tolman temperature
are also resolved.
\end{abstract}



\maketitle


\section{Introduction}
\label{sec:intro}
The proper temperature of the gravitating system of a perfect fluid
in thermodynamic equilibrium has been defined by the well-known Tolman temperature
\cite{Tolman:1930zza, Tolman:1930ona}.
In a static geometry, it assumes:
(i) the perfect fluid of radiation in thermal equilibrium,
(ii) the covariant conservation law of energy-momentum tensor,
(iii) the traceless condition of energy-momentum tensor,
(iv) the Stefan-Boltzmann law.
The resulting temperature
in the proper frame is written as
\begin{equation}\label{free}
 T_{{\rm T}} = \frac{C}{\sqrt{-g_{00}(r)}},
\end{equation}
where the Tolman factor appears in the denominator and $C$ is a constant determined by a boundary condition.
For example, for the Schwarzschild black hole, the constant used to be determined by $C=T_{\rm H}$,
 where $T_{\rm H}$ is the Hawking temperature of the black hole
\cite{Hawking:1974rv,Hawking:1974sw}.
As expected, the Tolman temperature becomes the Hawking temperature at infinity,
whereas it is infinite at the horizon due to the blue-shifted Tolman factor
which was  discussed in Ref. \cite{Frolov:2011}.
It is worth noting that the Tolman temperature is for the freely falling observer at rest
rather than the fixed observer who undergoes an acceleration \cite{Tolman:1930ona}.
For the fixed observer placed at the radius $r$ of the Schwarzschild black hole,
the temperature
can be expressed as the red/blue-shifted Hawking temperature
\begin{equation}\label{fixed}
 T_{{\rm F}} = \frac{T_H}{\sqrt{-g_{00}(r)}},
\end{equation}
where the red/blue-shift factor comes from the time dilation
in the presence of the gravitational field at different
places \cite{w}.
The fixed temperature is infinite at the horizon, which can also be understood in terms of the Unruh effect
for the large black hole
by keeping the detector in place \cite{Unruh:1976db}, since
the Unruh temperature is infinite at the horizon
because of the infinite acceleration of the frame.

First, it would be interesting to note that the two temperatures \eqref{free} and \eqref{fixed} are the same in spite of
the apparently different
physical backgrounds; the former is for the inertial frame and the latter is for the fixed one.
Second, the infinite Tolman temperature
at the horizon is much more puzzling unless $C=0$.
The firewall paradox was debated in evaporating black holes \cite{Almheiri:2012rt}
and a similar prediction based on different
assumptions was given in Ref.~\cite{Braunstein:2013bra}.
Note that this paradox can also be found even in the static black hole, since
the Tolman temperature \eqref{free} tells us that the freely falling observer encounters
quanta of the super-Planckian frequency at the horizon in the Hartle-Hawking-Israel state \cite{Hartle:1976tp, Israel:1976ur}.
The recent work for the firewall issue in
thermal equilibrium claims the existence of the massless firewall  \cite{Israel:2014eya} whose energy density is negligible but temperature
is infinite at the horizon.
Eventually, it leads to the violation of the equivalence principle at the horizon.

On the other hand, it was shown that the equivalence principle
can be restored at the horizon  by invoking that  at the horizon the
Unruh temperature measured by the accelerating detector is the same as the temperature measured by the fixed detector
in a gravitational field \cite{Singleton:2011vh}.
Additionally, in the Hartle-Hawking-Israel state,
the energy density and pressure are finite
at the horizon
even though the Tolman temperature is infinite at the horizon \cite{Page:1982fm,Visser:1996iw}.
It implies that
the Stefan-Boltzmann law
to relate the energy density (or the pressure) to the temperature must be nontrivial, which has been
unsolved yet.
Even worse, the energy density at the horizon is negative in the Hartle-Hawking-Israel state, so that
it seems to be nontrivial task to relate the negative energy density to the positive
temperature if the conventional
Stefan-Boltzmann law is just assumed.
In these respects, it raises some natural questions.
In spite of the finite energy density at the horizon,
what is the reason why the Tolman temperature is divergent at the horizon?
So, is there any consistent Stefan-Boltzmann law to relate the energy density to the
temperature covering the whole region?

In this work, we would like to investigate the Tolman temperature intensively
in order to resolve the above issues in the regime of the standard quantum field theory and thermodynamics.
To shed light on the essential feature of our formulation with exact solvability,
we adopt the two-dimensional approach to the problems.
First of all, we note that the energy-momentum tensor of matter fields on the classical background metric receives
semiclassical quantum corrections which give rise to the trace anomaly \cite{Deser:1976yx}.
It means that the Tolman relation \eqref{free} is correct for the traceless case; however, it should
be generalized semiclassically for a consistent formulation
when Hawking radiation is involved, since Hawking radiation is
indeed related to the trace anomaly of matter fields \cite{Christensen:1977jc}.
To get the consistent local proper temperature of the black hole,
the conditions (iii) and (iv) among the four assumptions in the original Tolman's derivation
should be released {\it ab initio}.

In Section \ref{sec:Tolman}, in the presence of the trace anomaly,
we will derive a trace anomaly-induced Stefan-Boltzmann law by
using the first law of thermodynamics and the nice property of the temperature independence of the trace anomaly
\cite{BoschiFilho:1991xz},
and naturally obtain the generalized Tolman temperature which can be reduced to
the conventional Tolman temperature if the traceless
condition is met. 
In Section \ref{sec:2D Sch}, for the exactly soluble two-dimensional Schwarzschild black hole,
we shall show that the generalized Tolman temperature becomes finite everywhere and it vanishes at the horizon
without the Tolman factor.
As a result, it will be shown that the equivalence principle survives
at the horizon thanks to the quantum principle, and the above-mentioned
questions in connection with the Tolman temperature
are also resolved.
Finally, conclusion and discussion will be given in Section \ref{sec:Diss}.

\section{Tolman temperature from trace anomaly-induced Stefan-Boltzmann law}
\label{sec:Tolman}
We start with a two-dimensional line element given as
\begin{equation}\label{metric}
ds^2=- f_1(r)dt^2+f_2(r) dr^2,
\end{equation}
where $f_1(r)$ and $f_2(r)$ are static functions and the metric is assumed to be asymptotically flat.
In the static system, the overall macroscopic velocity of radiation flow is zero, and
the velocity can be written as
\begin{equation}\label{velocity}
u^\mu=\frac{dx^\mu}{d\tau} = \left(\frac{1}{ \sqrt{f_1(r)}},~~0\right).
\end{equation}
The radiation is also regarded as a perfect fluid, so that
the energy-momentum tensor is written as
\begin{equation}\label{Tmunu}
T^{\mu\nu}=(\rho+p)u^\mu u^\nu +p g^{\mu\nu},
\end{equation}
where $\rho=T_{\mu\nu}u^\mu u^\nu$ and $p=T_{\mu \nu} n^\mu n^\nu$ are the local proper energy density and
pressure, respectively, and $n^\mu$ is the spacelike unit normal vector satisfying
$n^\mu n_\mu=1$ and $n^\mu u_\mu=0$.
Note that the flux is also calculated as ${\cal{F}}=-T_{\mu\nu}u^\mu n^\nu$ which is
 zero in the static fluid corresponding to the thermal radiation in equilibrium
\cite{Hartle:1976tp, Israel:1976ur}.
Next, the covariant conservation law of the energy-momentum tensor
can be written as
$2 f_1 \partial_r T^r_r = (T^t_t-T^r_r) \partial_r f_1$,
which is reduced to
\begin{equation}\label{source}
2 f_1 \partial_r p = -(\rho+p) \partial_r f_1.
\end{equation}
Next, the trace equation is given as
\begin{equation}\label{trace}
-\rho + p= T^\mu_\mu,
\end{equation}
where the trace of the energy-momentum tensor is not always zero.
Combining Eqs. \eqref{source} and \eqref{trace}, one can get
\begin{equation}\label{Dfp}
\partial_r (f_1 p)=\frac{1}{2}T^\mu_\mu \partial_r f_1.
\end{equation}
The resulting equation \eqref{Dfp} is easily solved as
\begin{equation}\label{improvedP}
p=\frac{1}{f_1}\left(C_0+\frac{1}{2}\int  T^\mu_\mu  df_1 \right),
\end{equation}
and
\begin{equation}\label{improvedrho}
\rho=\frac{1}{f_1}\left(C_0-f_1 T^\mu_\mu+\frac{1}{2}\int  T^\mu_\mu  df_1 \right),
\end{equation}
where the pressure and energy density are corrected by the trace anomaly, respectively.
Note that
the conventional Stefan-Boltzmann law in the two dimensional flat space is actually $p=\rho=\alpha T^2$ which is
valid only in the absence of the trace anomaly, where $\alpha$ is the Stefan-Boltzmann constant.
From Eqs. \eqref{improvedP} and \eqref{improvedrho}, the pressure and energy density are no longer
symmetric. Moreover, there are many different expressions
satisfying the anomaly relation \eqref{trace}.
To relate the pressure \eqref{improvedP} and energy density \eqref{improvedrho}
to the temperature uniquely, we should find the Stefan-Boltzmann law
which is compatible with the presence of the trace anomaly.

Now, for our purpose, the first law of thermodynamics is considered as
\begin{align}
dU =TdS-pdV,
\end{align}
where $U$, $T$, $S$, and $V$ are the thermodynamic internal energy, temperature, entropy, and
volume in the proper frame, respectively, and $U=\int \rho dV$.
Thus, the first law is rewritten in the form of
\begin{equation}
\left.\frac{\partial U}{\partial V} \right\vert_T  = T \left.\frac{\partial S}{\partial V}\right\vert_T-p.
\end{equation}
Using the Maxwell relation of $\partial S/\partial V \vert_T = \partial p/\partial T\vert_V$,
we get
\begin{equation}
\label{well}
\rho = T \left.\frac{\partial p}{\partial T}\right\vert_V-p.
\end{equation}
Next, we are going to use the fact that the trace anomaly is independent of the temperature \cite{BoschiFilho:1991xz},
so that from Eq. \eqref{trace}
we can obtain
\begin{equation}
\label{tt}
\left.\frac{\partial \rho}{\partial T}\right\vert_V =\left.\frac{\partial p}{\partial T}\right\vert_V,
\end{equation}
where $\partial_{T} T^\mu_\mu|_V =0$.
Plugging Eqs. \eqref{trace} and \eqref{tt} into Eq. \eqref{well} in order to eliminate the pressure and its derivative with
respect to the temperature with the fixed volume,
one can get the first order differential equation for the energy density given as
\begin{equation}
\label{key}
2\rho = T \left.\frac{\partial \rho}{\partial T}\right\vert_V - T^\mu_\mu.
\end{equation}
Solving Eq. \eqref{key}, the energy density and pressure can be obtained as
\begin{align}
\label{r}
\rho = \gamma T^2 - \frac{1}{2}T^\mu_\mu,
\end{align}
and
\begin{align}
\label{p}
p = \gamma T^2 + \frac{1}{2}T^\mu_\mu,
\end{align}
where they are reduced to the conventional ones for the traceless case if
the integration constant $\gamma$ is identified with the two-dimensional Stefan-Boltzmann constant, for example, $\gamma=\alpha = \pi/6$
for the massless scalar field \cite{Christensen:1977jc}.
Hence, from Eqs. \eqref{r} and \eqref{p}, the temperature can be written as
\begin{equation}
T=\sqrt{\frac{1}{\alpha}\left(p-\frac{1}{2} T^\mu_\mu \right)} =\sqrt{\frac{1}{\alpha}\left(\rho +\frac{1}{2} T^\mu_\mu \right)}.
\end{equation}
Therefore, the resulting generalized Tolman temperature by using Eqs. \eqref{improvedP} or \eqref{improvedrho} is obtained as
\begin{equation}\label{improvedT}
T=\frac{1}{\sqrt{\alpha f_1}} \sqrt{C_0 - \frac{f_1}{2}T^\mu_\mu+\frac{1}{2}\int  T^\mu_\mu  df_1 },
\end{equation}
where the temperature is independent of $f_2$.
Indeed, there appeared nontrivial contributions to the temperature from the trace anomaly.
Note that it is reduced to the conventional Tolman temperature if the energy-momentum tensor is traceless, so that
$T = C/\sqrt{f_1(r)}$,
where $C =\sqrt{C_0/\alpha}$.
In the asymptotic infinity, the trace parts in Eq. \eqref{improvedT} vanish,
and the constant $C_0$ can be determined by the usual boundary condition.

\section{Application to two-dimensional Schwarzschild black hole}
\label{sec:2D Sch}

Let us now study how the generalized Tolman temperature
\eqref{improvedT} actually works in the two-dimensional Schwarzschild black hole, where
the metric is given as
\begin{equation}
\label{metric1}
f(r) = f_1(r) = \frac{1}{f_2(r)} = 1-\frac{2M}{r},
\end{equation}
where $M$ is the mass of black hole and the Newton constant is set to $G=1$.
Now, using the explicit trace anomaly for the massless scalar field as
$T^\mu_\mu = R/(24\pi)$ \cite{Deser:1976yx,Christensen:1977jc},
the proper temperature \eqref{improvedT} can be calculated
by imposing the boundary condition of $C_0 = \alpha/(8\pi M)^2$
which gives the Hawking temperature at infinity,
\begin{align}
\label{tttt}
T=\frac{1}{8\pi M \sqrt{f(r)}} \sqrt{ 1 - 4\left(\frac{2M}{r}\right)^3+3\left(\frac{2M}{r}\right)^4}.
\end{align}
The quantities in the square root in Eq. \eqref{tttt} can be factorized as
\begin{align}
T=&\frac{1}{8\pi M \sqrt{f(r)}} \notag \\
&\times \sqrt{ \left(1 - \frac{2M}{r}\right) \left(1+ \frac{2M}{r}+\left( \frac{2M}{r}\right)^2-3\left( \frac{2M}{r}\right)^3 \right)},
\end{align}
and consequently the generalized Tolman temperature is obtained as
\begin{equation}\label{newT}
T=\frac{1}{8\pi M}\sqrt{1+ \frac{2M}{r}+\left( \frac{2M}{r}\right)^2-3\left( \frac{2M}{r}\right)^3}.
\end{equation}
\begin{figure}[pt]
  \begin{center}
  \includegraphics[width=0.45\textwidth]{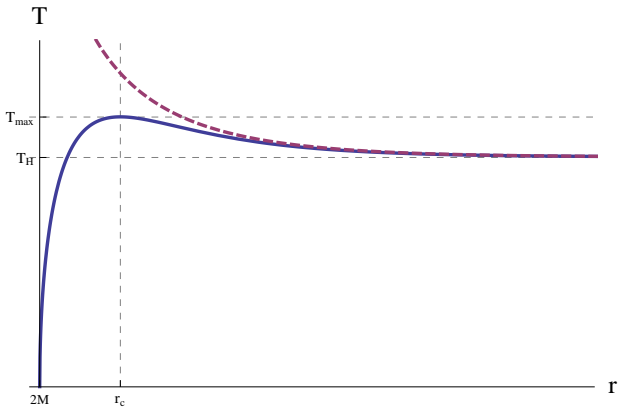}
  \end{center}
  \caption{The thick dotted curve is for the conventional Tolman temperature which is infinite at the horizon, whereas
  the solid curve is for the generalized one which is finite everywhere and especially goes to zero at the horizon.
  The maximum of the latter temperature $T_{\rm max}$
  occurs at $r_c \sim 4M$ in our model. The constant is set to
  $M=1$ for convenience. The infinite Tolman temperature
  at the horizon was suppressed by taking into account the trace anomaly.}
  \label{fig:Tolman}
\end{figure}
Note that the Tolman factor does not appear, which is compared to the form of the conventional Tolman temperature
\eqref{free}.
One of the most interesting things to distinguish from the conventional behaviors of the Tolman temperature
is that it is finite everywhere, and it also has a maximum value of
the temperature at $r_c \sim 4M $ as seen from Fig. \ref{fig:Tolman}.
In particular, the temperature vanishes at the horizon.
The suppression of the infinite Tolman temperature at the horizon by means of the quantum-mechanical trace anomaly
is reminiscent of that of the infinite intensity at the high frequency in Rayleigh-Jeans law
by the quantum correction.

As a matter of fact, in the large black hole, the metric \eqref{metric1}
could be described by the Rindler metric for the near horizon limit,
and the Unruh effect tells us that the temperature is given as $T_{\rm U}=a/2\pi$
in terms of the proper acceleration, where the acceleration of the fixed frame is $a= M/( r^2 \sqrt{f(r)})$
\cite{Unruh:1976db}.
It implies that the free-fall observer would find the vanishing Unruh temperature,
if the frame were free from the acceleration.
So, it is reasonable for the observer in the proper frame
to get the vanishing temperature at the horizon rather than
the infinite temperature.
In addition to this, authors  in Ref. \cite{Singleton:2011vh} also showed that
the temperature \eqref{fixed} measured by the fixed observer in
the gravitational background is generically higher than the Unruh temperature of the accelerating observer;
however,
they are the same at the event horizon of the black hole, so that the equivalence principle in the
quantized theory is restored at the horizon.
Thus, the vanishing generalized Tolman temperature at the horizon is
compatible with the result that the equivalence principle
is recovered at the horizon in Ref. \cite{Singleton:2011vh}.

Let us make a comment on the energy density and pressure.
Plugging the generalized Tolman temperature \eqref{newT} into Eqs. \eqref{r} and \eqref{p},
one can obtain
\begin{align}
\rho &= - \frac{1}{48\pi r^4 f(r)} \left(8Mr f(r)+2M^2-\frac{r^4}{8M^2}\right) \label{rho} , \\
p &=\frac{1}{384 \pi M^2} \left[1+ \frac{2M}{r}+\left( \frac{2M}{r}\right)^2+\left( \frac{2M}{r}\right)^3\right] \label{pp},
\end{align}
where the energy density and pressure near the horizon are negative and positive finite as
$-\rho = p=1/(96\pi M^2)$, while $\rho=p = 1/(384\pi M^2)=(\pi/6 )T_H^2$ at infinity.
In a self-contained manner, let us confirm whether the above energy density and pressure calculated by
employing the generalized Tolman temperature
are consistent with the results from direct calculations or not.
For this purpose, in the light-cone coordinates defined as $\sigma^{\pm}=t \pm r^*$ through $r^*=r+2M \ln(r/M-2)$,
the proper velocity \eqref{velocity} can be written as $u^+ =u^- =1 /\sqrt{f(r)}$,
where  $u^\pm=u^t \pm u^r/f(r)$ and
$ n^+ =-n^- =1 /\sqrt{f(r)}$.
The components of the energy-momentum tensor are expressed as
$T_{\pm \pm} = -({1}/{48 \pi}) ({2M
 f(r)}/{r^3} + {M^2}/{r^4} ) + ({1}/{48})t_{\pm}$ and
   $T_{+-}  = -({1}/{48\pi} ) ({2M}f(r)/r^3)$, where $t_{\pm} $ are the integration functions
   obtained from the integration of the covariant conservation law.
The energy density and pressure measured in the freely falling frame can be
calculated as
$  \rho=  -1/(48 \pi r^4 f(r)) [8Mr f(r)+2M^2 -\pi r^4(t_++t_-) ] $ \cite{Eune:2014eka} and
  $ p=   1/(48 \pi r^4 f(r)) [-2M^2 +\pi r^4(t_++t_-) ]$,
where we used the definition for the freely falling energy density and pressure.
Since the radiation flow in the Hartle-Hawking-Israel state is characterized by choosing the integration functions as
$t_\pm=1/(16\pi M^2)$ \cite{Wipf:1998ss}, one can easily see that
 Eqs. \eqref{rho} and \eqref{pp} derived from the generalized Tolman temperature
\eqref{newT}
are coincident with the above energy density and pressure
based on the standard calculations.


\section{Discussion and Conclusion}
\label{sec:Diss}

It would be interesting to compare our computations with
a previous result. The temperature \eqref{newT} looks different from the free-fall temperature at rest,
  $T_{\rm BT}(r)=(1/8\pi M)$
  $\sqrt{1+ 2M/r+( 2M/r)^2+(2M/r)^3}$ \cite{Brynjolfsson:2008uc}
calculated by using the global embedding of the four-dimensional Schwarzschild black hole into
a higher dimensional flat spacetime \cite{Deser:1998xb}.
For example, the value of $T_{\rm BT}$ at the horizon is larger than that of the temperature at infinity,
precisely, $T_{\rm BT}(2M)=2T_{\rm H}$ which is a maximum.
Simply, we cannot conclude that the difference between them comes from the dimensionality, since
we can exactly get the same free-fall temperature as $T_{\rm BT}$ for
the two-dimensional Schwarzschild black hole \eqref{metric1} by using
a slight different higher-dimensional embedding method \cite{Banerjee:2010ma}.
Instead, we consider the new expression for the Stefan-Boltzmann law such as
$p=\alpha T^2$ and
$\rho= \alpha T^2 - T^\mu_\mu$, then
$T_{\rm BT}$ can be obtained from Eq. \eqref{pp}; however,
this does not satisfy the relation \eqref{well} which comes from the first law of thermodynamics.
Therefore, if the first law of thermodynamics  is valid in the proper frame,
 the unique Stefan-Boltzmann law can be obtained thermodynamically among diverse expressions to satisfy
the anomaly equation \eqref{trace}.

For the massless firewall in Ref. \cite{Israel:2014eya},
it was claimed that it is massless but hot in the Hartle-Hawking-Israel state of black holes.
At first sight, this phenomenon seems to be plausible in that
the energy density and pressure at the horizon are at most negligible order of
$1/M^2$ in comparison with
that of the temperature. Moreover, the infinite Tolman temperature at the horizon
indicates the existence of the hot object.
However, employing the generalized temperature \eqref{newT},
one could evade the infinite temperature at the horizon, and thus
save the violation of the equivalence principle.

In conclusion,
we have shown that the conventional Tolman temperature derived from the
assumption of the traceless condition of energy-momentum tensor
for matter fields was generalized, since
the temperature associated with Hawking radiation is related to the trace anomaly.
The most important ingredient in our formulation is that the Stefan-Boltzmann law
was generalized in the presence of the trace anomaly
by using the first law of thermodynamics and the property of the temperature independence of the
trace anomaly. As a result, we obtained the generalized Tolman temperature
 which can be reduced to the conventional Tolman temperature if the traceless condition is met.
In terms of the two-dimensional Schwarzschild black hole,
we showed that
the generalized Tolman temperature becomes finite everywhere and,
in particular, it vanishes at the horizon, while it approaches the Hawking temperature at infinity.
The quantum principle does not always give rise to conflicts
but sometimes plays a key role to maintaining the equivalence principle.
We hope that such a modification of the temperature as Eq. \eqref{improvedT}
will provide some clues about paradoxical problems in quantum gravity.
Finally, it would be
interesting to generalize this approach to higher dimensional black holes
 and other gravitational models, including some classical models whose traces
are nontrivial.

\acknowledgments
We would like to thank M. S. Eune  and Edwin J. Son for exciting discussions.
This work was supported by the National Research Foundation of Korea(NRF) grant funded by the Korea government(MSIP) (2014R1A2A1A11049571).


\end{document}